\documentclass[pre,twocolumn,superscriptaddress,nofootinbib]{revtex4}
\usepackage{graphicx,chemarr}

\begin{document}

\title{Temporal precision of regulated gene expression}

\author{Shivam Gupta}
\affiliation{Department of Physics and Astronomy, Purdue University, West Lafayette, Indiana 47907, USA}

\author{Julien Varennes}
\affiliation{Department of Physics and Astronomy, Purdue University, West Lafayette, Indiana 47907, USA}

\author{Hendrik C.\ Korswagen}
\affiliation{Hubrecht Institute, Royal Netherlands Academy of Arts and Sciences and University Medical Center Utrecht, Uppsalalaan 8, 3584 CT Utrecht, the Netherlands}

\author{Andrew Mugler}
\email{amugler@purdue.edu}
\affiliation{Department of Physics and Astronomy, Purdue University, West Lafayette, Indiana 47907, USA}

\begin{abstract}
Important cellular processes such as migration, differentiation, and development often rely on precise timing. Yet, the molecular machinery that regulates timing is inherently noisy. How do cells achieve precise timing with noisy components? We investigate this question using a first-passage-time approach, for an event triggered by a molecule that crosses an abundance threshold and that is regulated by either an accumulating activator or a diminishing repressor. We find that the optimal strategy corresponds to a nonlinear increase in the amount of the target molecule over time. Optimality arises from a tradeoff between minimizing the extrinsic timing noise of the regulator, and minimizing the intrinsic timing noise of the target molecule itself. Although either activation or repression outperforms an unregulated strategy, when we consider the effects of cell division, we find that repression outperforms activation if division occurs late in the process. Our results explain the nonlinear increase and low noise of {\it mig-1} gene expression in migrating neuroblast cells during {\it Caenorhabditis elegans} development, and suggest that {\it mig-1} regulation is dominated by repression for maximal temporal precision. These findings suggest that dynamic regulation may be a simple and powerful strategy for precise cellular timing.
\end{abstract}

\maketitle

Proper timing is crucial for biological processes, including cell division \cite{bean2006coherence, nachman2007dissecting, schneider2004growth}, cell differentiation \cite{carniol2004threshold}, cell migration \cite{mentink2014cell}, viral infection \cite{amir2007noise}, embryonic development \cite{meinhardt1982models, tufcea2015critical}, and cell death \cite{roux2015fractional}. These processes are governed by molecular events inside cells, i.e., production, degradation, and interaction of molecules. Molecular events are subject to unavoidable fluctuations, because molecule numbers are small and reactions occur at random times \cite{van1992stochastic, mcadams1997stochastic}. Cells combat these fluctuations using networks of regulatory interactions among molecular species. This raises the fundamental question of whether there exist regulatory strategies that maximize the temporal precision of molecular events and, in turn, cellular behaviors.

A canonical mechanism by which a molecular event triggers a cellular behavior is accumulation to a threshold \cite{ghusinga2017first, co2017stochastic, yurkovsky2012event, schneider2004growth, carniol2004threshold}: molecules are steadily produced by the cell, and once the molecule number crosses a particular threshold, the behavior is initiated. The temporal precision of the behavior is therefore bounded by the temporal precision of the threshold crossing. Threshold crossing has been shown to underlie cell cycle progression \cite{schneider2004growth} and sporulation \cite{carniol2004threshold}, although alternative strategies, such as derivative \cite{roux2015fractional} or integral thresholding \cite{jafarpour2017biological}, oscillation \cite{monti2016accuracy}, and dynamical transitions in the regulatory network \cite{tufcea2015critical}, have also been investigated.

Recent work has investigated the impact of auto-regulation (i.e., feedback) on the temporal precision of threshold crossing \cite{ghusinga2017first, co2017stochastic}. Interestingly, it was found that auto-regulation generically decreases the temporal precision of threshold crossing, meaning that the optimal strategy is a linear increase of the molecule number over time with no auto-regulation \cite{ghusinga2017first} (although auto-regulation can help if there is a large timescale separation and the threshold itself is also subject to optimization \cite{co2017stochastic}). Indeed, even when the molecule also degrades, the optimal precision is achieved when positive auto-regulation counteracts the effect of degradation, preserving the linear increase over time \cite{ghusinga2017first}. However, in many biological processes, such as the temporal control of neuroblast migration in {\it Caenorhabditis elegans} \cite{mentink2014cell}, the molecular species governing the behavior increases nonlinearly over time. This suggests that other regulatory interactions beyond auto-regulation may play an important role in determining temporal precision. In particular, the impact of activation and repression on temporal precision, where the activator or repressor has its own stochastic dynamics, remains unclear.

Here we investigate the temporal precision of threshold crossing for a molecule that is regulated by either an accumulating activator or a degrading repressor. Using a first-passage-time approach \cite{ghusinga2017first, redner2001guide, chou2014first, iyer2016first} and a combination of computational and analytic methods, we find that, unlike in the case of auto-regulation, the presence of either an activator or a repressor increases the temporal precision beyond that of the unregulated case. Furthermore, the optimal regulatory strategy for either an activator or a repressor corresponds to a nonlinear increase in the regulated molecule number over time. We elucidate the physical mechanism behind these optimal strategies, which stems from a tradeoff between reducing the noise of the regulator and reducing the noise of the target molecule. Motivated by data from migrating neuroblast cells in {\it C.\ elegans} larvae \cite{mentink2014cell}, we also consider the effects of cell division, and find that activation (repression) is optimal if cell division occurs early (late) in the temporal process. Our results are quantitatively consistent with both the temporal precision and nonlinearity of the {\it mig-1} mRNA dynamics in the neuroblasts, and we predict that {\it mig-1} regulation is dominated by repression for maximal temporal precision. The agreement of our simple model with these data suggests that many molecular timing processes may benefit from the generic regulatory strategies we identify here.

\section{Results}

\begin{figure}
\begin{center}
\includegraphics[width=.5\textwidth]{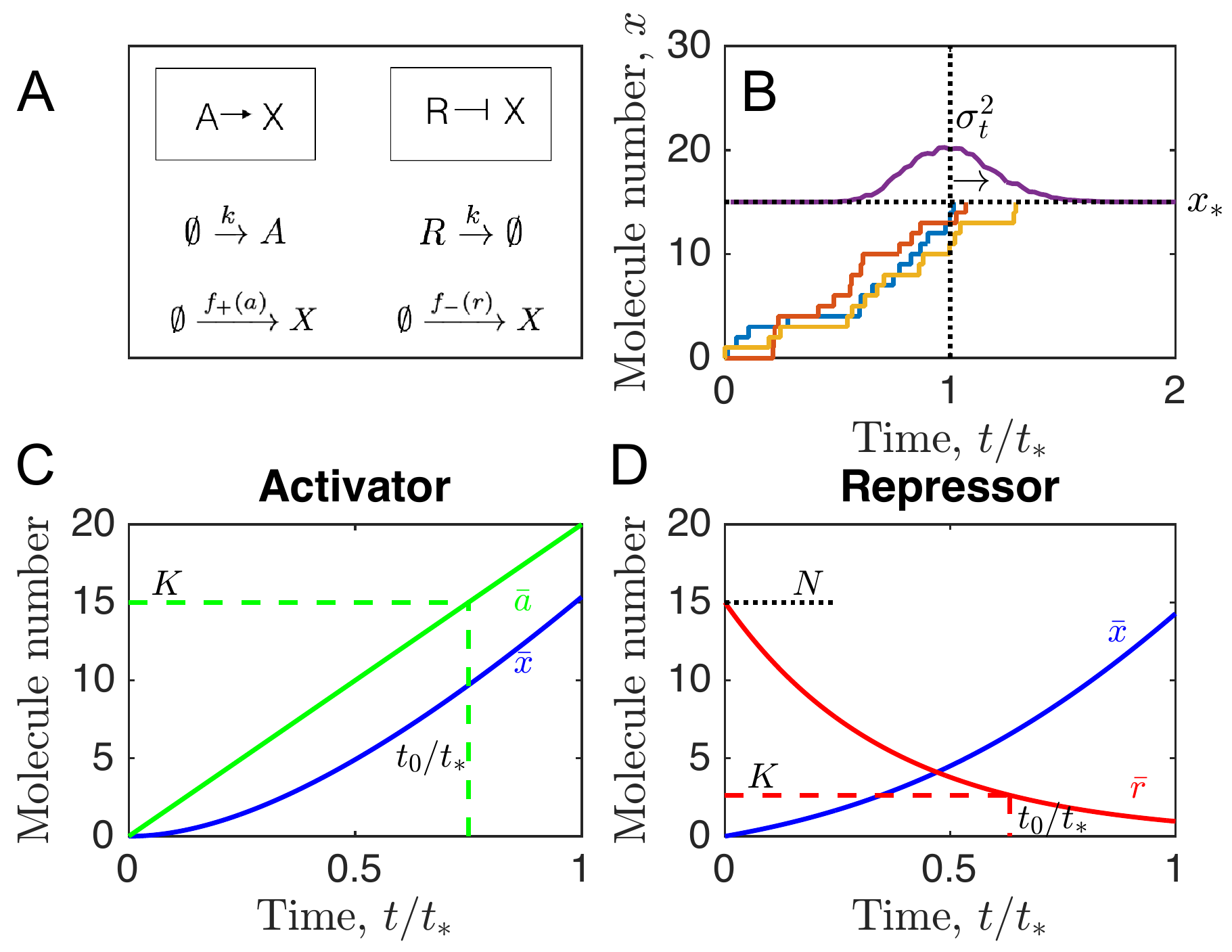}
\end{center}
\caption{Threshold crossing of a regulated molecular species. (A) A target species $X$ is regulated by either an accumulating activator $A$ or a degrading repressor $R$. (B) Temporal precision is quantified by the variance $\sigma_t^2$ of the first-passage time, at which the stochastic molecule number $x$ first crosses the threshold $x_*$. (C, D) Deterministic dynamics illustrate the effects of regulation. Parameters are $kt_* = 20$ and $K = 15$ in C; $kt_* = 2.75$, $K = 2.6$, and $N = 15$ in B and D; and $x_* = 15$ and $H = 1$ throughout. $t_0$ is defined by $\bar{a}(t_0) = K$ in C and $\bar{r}(t_0) = K$ in D.}
\label{setup}
\end{figure}

We consider a molecular species $X$ whose production is regulated by a second species, either an activator $A$ or a repressor $R$ (Fig.\ \ref{setup}A). The regulator undergoes its own dynamics: the activator is produced at a rate $k$ whereas the repressor is degraded at a rate $k$, such that in either case the production rate of $X$ increases over time. For the regulation function we take a Hill function, which is a generic model of cooperative regulation \cite{walczak2012analytic, ghusinga2017first, co2017stochastic},
\begin{align}
\label{f+}
f_+(a) = \frac{\alpha a^H}{a^H+K^H} \quad {\rm (activator)}, \\
\label{f-}
f_-(r) = \frac{\alpha K^H}{r^H+K^H} \quad {\rm (repressor)}.
\end{align}
Here $a$ and $r$ are the molecule numbers of $A$ and $R$, respectively, $\alpha$ is the maximal production rate of $X$, $K$ is the half-maximal regulator number, and $H$ is the cooperativity.

We suppose that a behavior is initiated when the molecule number $x$ crosses a threshold $x_*$ (Fig.\ \ref{setup}B). Because the production of $X$ and the dynamics of the regulator are stochastic, the time at which $x$ first reaches $x_*$ is a random variable. We characterize the precision of this event by the mean $\bar{t}$ and variance $\sigma_t^2$ of this first-passage time, which we compute numerically from the master equation corresponding to the reactions in Fig.\ \ref{setup}A (see Materials and Methods). The maximal production rate $\alpha$ is set to ensure that $\bar{t}$ is equal to a target time $t_*$, which we assume is set by functional constraints on the initiated behavior. This leaves $k$, $K$, and $H$ as free parameters of the regulation that can, in principle, be optimized to minimize the timing variance $\sigma_t^2$.

The deterministic dynamics, illustrated in Fig.\ \ref{setup}C and D, neglect fluctuations but give an intuitive picture of the regulation. Whereas the amount of activator increases linearly over time, the amount of repressor decays exponentially from an initial molecule number $N$:
\begin{align}
\label{abar}
\bar{a}(t) &= kt, \\
\label{rbar}
\bar{r}(t) &= Ne^{-kt}.
\end{align}
In either case, the production rate $f_\pm$ of $X$ increases over time, such that $\bar{x}$ increases nonlinearly. $N$ is an additional free parameter in the repressor case.

\subsection{Regulation increases temporal precision}

To investigate the effects of regulation on temporal precision, we consider the timing variance $\sigma_t^2$ as a function of the parameters $k$ and $K$. The special case of no regulation corresponds to the limits $k\to\infty$ and $K\to0$ in the case of activation, or $k\to\infty$ and $K\to\infty$ in the case of repression. In this case, the production of $X$ occurs at the constant rate $\alpha$. Reaching the threshold requires $x_*$ sequential events, each of which occurs in a time that is exponentially distributed with mean $1/\alpha$. The total completion time for such a process is given by a gamma distribution with mean $\bar{t} = x_*/\alpha$ and variance $\sigma_t^2 = x_*/\alpha^2$ \cite{iyer2016first}. Ensuring that $\bar{t} = t_*$ requires $\alpha = x_*/t_*$, for which the variance satisfies $\sigma_t^2x_*/t_*^2 = 1$. This expression gives the timing variance for the unregulated process.

In Fig.\ \ref{heatmaps} we plot the scaled variance $\sigma_t^2x_*/t_*^2$ as a function of the parameters $k$ and $K$ for cooperativity $H = 1$ (color maps). In the case of activation (Fig.\ \ref{heatmaps}A), the variance decreases with increasing $k$ and $K$. This means that the temporal precision is highest for an activator that accumulates quickly and requires a high abundance to produce $X$. In the case of repression (Fig.\ \ref{heatmaps}B), the variance has a global minimum as a function $k$ and $K$. This means that the temporal precision is highest for a repressor with a particular well-defined degradation rate and abundance threshold. Importantly, we see that for both activation and repression, the scaled variance can be less than one, meaning that regulation allows improvement of the temporal precision beyond that of the unregulated process.

\subsection{Optimal regulation balances intrinsic and extrinsic noise}

\begin{figure}
\begin{center}
\includegraphics[width=.5\textwidth]{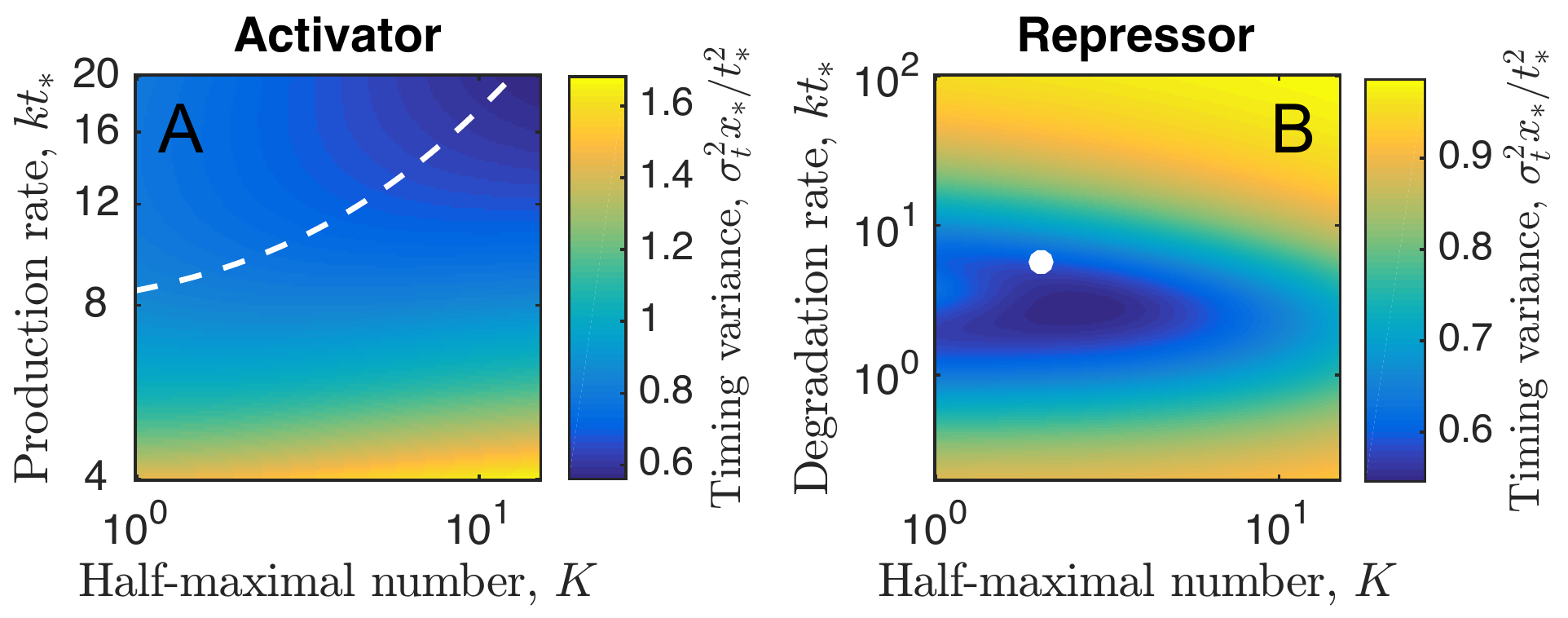}
\end{center}
\caption{Optimal regulatory strategies. Timing variance as a function of the regulatory parameters reveals (A) a trajectory along which the variance decreases in the case of the activator and (B) a global minimum in the case of the repressor. White dashed line in A and white dot in B show the analytic approximations in Eqs.\ \ref{minA} and \ref{minR}, respectively. Parameters are $N = 15$ in B, and $x_* = 15$ and $H = 1$ in both.}
\label{heatmaps}
\end{figure}

To understand the dependencies in Fig.\ \ref{heatmaps}, we develop analytic approximations. First, we assume that $H\to\infty$, such that the regulation functions in Eqs.\ \ref{f+} and \ref{f-} become threshold functions. In this limit, the production rate of $X$ is zero if $a<K$ or $r>K$, and $\alpha$ otherwise. The deterministic dynamics of $X$ become piecewise-linear,
\begin{equation}
\label{xbar}
\bar{x}(t) =
\begin{cases}
0 & t < t_0 \\
\alpha (t-t_0) & t \ge t_0,
\end{cases}
\end{equation}
where $t_0$ is determined by either $\bar{a}(t_0) = K$ or $\bar{r}(t_0) = K$ according to Eqs.\ \ref{abar} and \ref{rbar}.
Then, to set $\alpha$, we use the condition $\bar{x}(t_*) = x_*$, which results in $\alpha = x_*/(t_*-t_0)$.

Lastly, we approximate the variance in the first-passage time using the variance in the molecule number and the time derivative of the mean dynamics \cite{co2017stochastic}. Specifically, the timing variance of $X$ arises from two sources: (i) uncertainty in the time when the regulator crosses its threshold $K$, which determines when the production of $X$ begins, and (ii) uncertainty in the time when $x$ crosses its threshold $x_*$, given that production begins at a particular time. The first source is extrinsic noise, and the second source is intrinsic noise. We estimate these timing variances from the associated molecule number variances, propagated via the time derivatives,
\begin{equation}
\label{ie}
\sigma^2_t \approx
	\underbrace{\left.\sigma_y^2\left(\frac{d\bar{y}}{dt}\right)^{-2}\right|_{t_0}}_{\rm extrinsic}
	+ \underbrace{\left.\sigma_x^2\left(\frac{d\bar{x}}{dt}\right)^{-2}\right|_{t_*}}_{\rm intrinsic},
\end{equation}
where $y \in \{a,r\}$ denotes the regulator molecule number.

For the activator, which undergoes a pure production process with rate $k$, the molecule number obeys a Poisson distribution with mean $kt$. Therefore, the molecule number variance at time $t_0$ is $\sigma_a^2 = kt_0$. For the repressor, which undergoes a pure degradation process with rate $k$ starting from $N$ molecules, the molecule number obeys a binomial distribution with number of trials $N$ and success probability $e^{-kt}$. Therefore, the molecule number variance at time $t_0$ is $\sigma_r^2 = Ne^{-kt_0}(1-e^{-kt_0})$. For the target molecule, which undergoes a pure production process with rate $\alpha$ starting at time $t_0$, the molecule number obeys a Poisson distribution with mean $\alpha(t-t_0)$. Therefore, the molecule number variance at time $t_*$ is $\sigma_x^2 = \alpha(t_*-t_0)$. Inserting these expressions into Eq.\ \ref{ie}, along with the derivatives calculated from Eqs.\ \ref{abar}-\ref{xbar} and the appropriate expressions for $\alpha$ and $t_0$, we obtain
\begin{align}
\label{varA}
\frac{\sigma^2_tx_*}{t_*^2} &\approx \frac{Kx_*}{(kt_*)^2} + \left(1 - \frac{K}{kt_*}\right)^2 \quad {\rm (activator)}, \\
\label{varR}
\frac{\sigma^2_tx_*}{t_*^2} &\approx \frac{(N-K)x_*}{NK(kt_*)^2} + \left[1 - \frac{\log(N/K)}{kt_*} \right]^2 \quad {\rm (repressor)}.
\end{align}
As a function of $kt_*$ and $K$, the global minimum of Eq.\ \ref{varA} occurs as $kt_*\to\infty$ and $K\to\infty$. The approach to this minimum is given by differentiating with respect to $K$ at fixed $kt_*$ and setting the result to zero, which yields the trajectory
\begin{equation}
\label{minA}
K  = 
\begin{cases}
0 & kt_* < \frac{x_*}{2} \\
kt_* - \frac{x_*}{2} & kt_* \ge \frac{x_*}{2},
\end{cases}
\end{equation}
along which the variance satisfies
\begin{equation}
\label{sigA}
\frac{\sigma^2_tx_*}{t_*^2} =
\begin{cases}
1 & kt_* < \frac{x_*}{2} \\
\frac{x_*}{kt_*}\left(1 - \frac{x_*}{4kt_*}\right) & kt_* \ge \frac{x_*}{2}.
\end{cases}
\end{equation}
In contrast, the global minimum of Eq.\ \ref{varR} occurs at finite $kt_*$ and $K$: differentiating with respect to each and setting the results to zero gives the values
\begin{subequations}
\label{minR}
\begin{align}
\label{minR1}
K &= e^{-2}N, \\
\label{minR2}
kt_* &= \frac{e^2x_*}{2N} + 2,
\end{align}
\end{subequations}
\begin{equation}
\label{sigR}
\frac{\sigma^2_tx_*}{t_*^2} = \frac{x_*}{x_*+4e^{-2}N},
\end{equation}
where we have assumed that $K/N \ll 1$ (see Materials and Methods), which is justified post-hoc by Eq.\ \ref{minR1}.

These analytic approximations are compared with the numerical results for the activator in Fig.\ \ref{heatmaps}A (white dashed line, Eq.\ \ref{minA}) and for the repressor in Fig.\ \ref{heatmaps}B (white circle, Eq.\ \ref{minR}). In Fig.\ \ref{heatmaps}A we see that the global minimum indeed occurs as $kt_*\to\infty$ and $K\to\infty$, and the predicted trajectory agrees well with the observed descent. In Fig.\ \ref{heatmaps}B we see that the predicted global minimum lies very close to the observed global minimum. The success of these approximations is especially striking given that the numerics are shown for $H = 1$, whereas the approximations take $H\to\infty$.

The success of the approximations means that Eq.\ \ref{ie} describes the key mechanism leading to the optimal temporal precision. Eq.\ \ref{ie} demonstrates that the optimal regulatory strategy arises from a tradeoff between minimizing extrinsic and intrinsic noise. On the one hand, minimizing only the extrinsic noise would require that the regulator cross its threshold $K$ very soon and with a steep slope, meaning that the target molecule would be effectively unregulated and would increase linearly over time. On the other hand, minimizing only the intrinsic noise would require that the regulator cross its threshold only shortly before the target time $t_*$, such that the target molecule would cross its threshold $x_*$ with a steep slope, leading to a highly nonlinear increase of the target molecule over time. In actuality, the optimal strategy is somewhere in between, with the regulator crossing its threshold at some intermediate time $t_0$, and the target molecule exhibiting moderately nonlinear dynamics as in Fig.\ \ref{setup}C and D.

Eqs.\ \ref{sigA} and \ref{sigR} demonstrate that the timing variance is small for large $kt_*/x_*$ in the case of activation, and small for large $N/x_*$ in the case of repression. This makes intuitive sense because each of these quantities scales with the number of regulator molecules: $k$ is the production rate of activator molecules, while $N$ is the initial number of repressor molecules. To make this intuition quantitative, we define a cost as the time-averaged number of regulator molecules,
\begin{align}
\label{costA}
\langle a\rangle &= \int_0^{t_*}dt\ \bar{a}(t) = \frac{1}{2}kt_*, \\
\label{costR}
\langle r\rangle &= \int_0^{t_*}dt\ \bar{r}(t) = \frac{N}{kt_*}(1-e^{-kt_*}),
\end{align}
where the second steps follow from Eqs.\ \ref{abar} and \ref{rbar}. We see that, indeed, $\langle a\rangle$ scales with $k$, and $\langle r\rangle$ scales with $N$. Thus, Eqs.\ \ref{sigA} and \ref{sigR} demonstrate that increased temporal precision comes at a cost, in terms of the number of regulator molecules that must be produced.

\subsection{Model predictions are consistent with neuroblast migration data}

\begin{figure}
\begin{center}
\includegraphics[width=.5\textwidth]{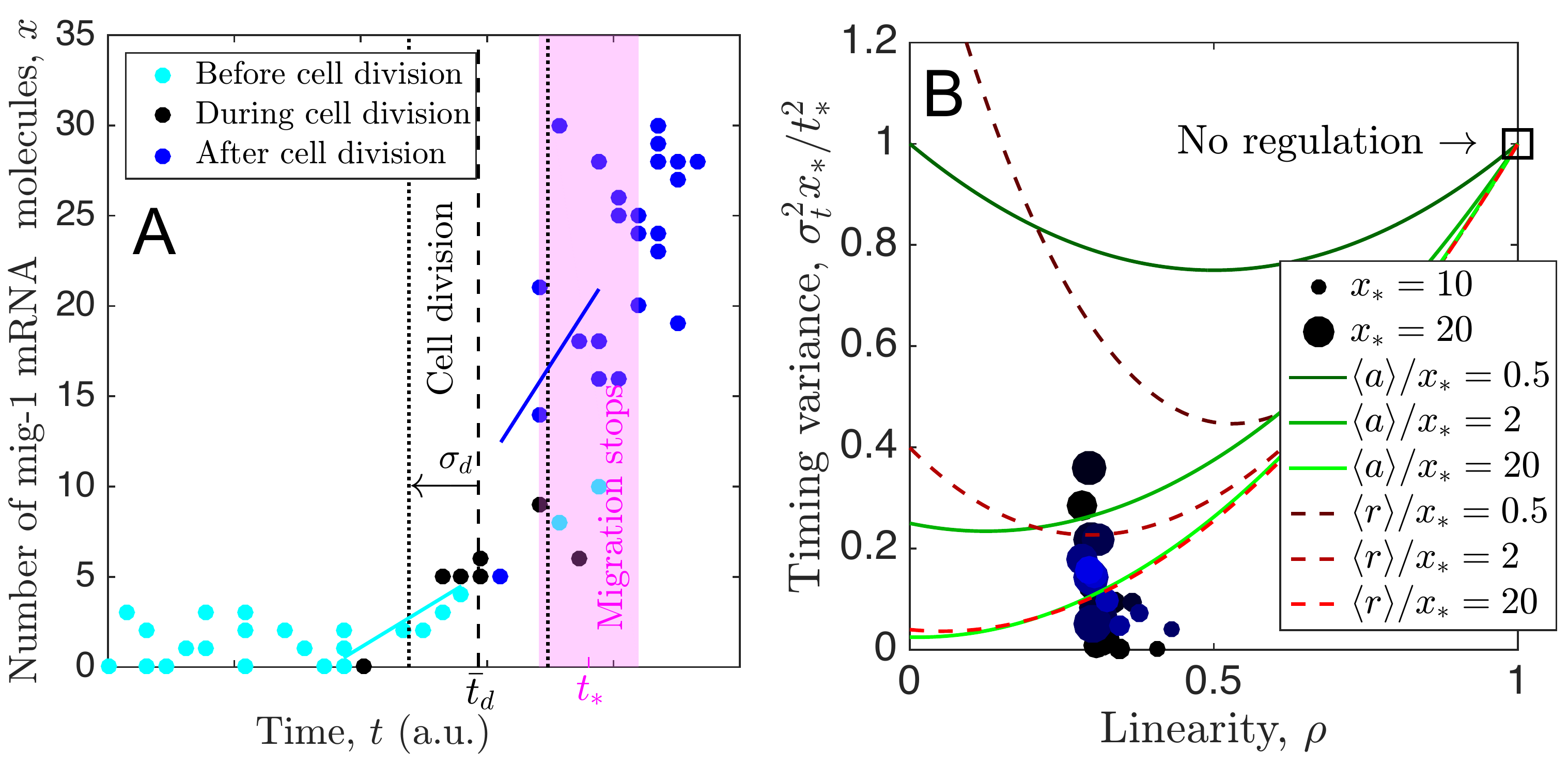}
\end{center}
\caption{Model predictions agree with neuroblast migration data. (A) Number of {\it mig-1} mRNA molecules per cell as a function of time $t$, obtained by single-molecule fluorescent in situ hybridization, from \cite{mentink2014cell}. Magenta shows approximate range of times when cell migration terminates. Black lines show mean $\bar{t}_d$ (dashed) and standard deviation $\sigma_d$ of cell division times (black points). Light and dark blue lines show fits to $D=8$ points closest to $\bar{t}_d - \sigma_d$ and $\bar{t}_d + \sigma_d$, respectively. (B) Timing variance vs.\ linearity of $x(t)$, both for experimental data in A (circles) and our model (curves, Eqs.\ \ref{rhoA} and \ref{rhoR}). Data analyzed using ranges of threshold $10 \le x_* \le 25$ (size) and bin size $3 \le \Delta x\le 12$ (color, from black to blue). We see that for sufficiently large cost $\langle a\rangle/x_*$ or $\langle r\rangle/x_*$, model predictions agree with experimental data.}
\label{expt}
\end{figure}

We test our model predictions using data from neuroblast cells in {\it C.\ elegans} larvae \cite{mentink2014cell}. During {\it C.\ elegans} development, particular neuroblast cells migrate from the posterior to the anterior of the larva. It has been shown that the migration terminates not at a particular position, but rather after a particular amount of time, and that the termination time is controlled by a temporal increase in the expression of the {\it mig-1} gene \cite{mentink2014cell}. Since {\it mig-1} is known to be subject to regulation \cite{ji2013feedback}, we investigate the extent to which the dynamics of {\it mig-1} can be explained by the predictions of our model.

Figure \ref{expt}A shows the number $x$ of {\it mig-1} mRNA molecules per cell as a function of time $t$, obtained by single-molecule fluorescent in situ hybridization (from \cite{mentink2014cell}). We analyze these data in the following way (see Materials and Methods for details). We see that the dynamics are nonlinear, and therefore we quantify the linearity using the area under the curve, normalized by that for a perfectly linear trajectory,
\begin{equation}
\label{rho}
\rho = \frac{2}{x_*t_*}\int_0^{t_*} dt\ x(t).
\end{equation}
By this definition, $\rho = 1$ for perfectly linear dynamics, and $\rho\to0$ for maximally nonlinear dynamics (a sharp rise at $t_*$). We estimate $x_*$, $t_*$, and the timing variance $\sigma_t^2$ from the data. Specifically, migration is known to terminate between particular reference cells in the larva \cite{mentink2014cell}, which gives an estimated range for the termination time $t_*$. This range is shown in magenta in Fig.\ \ref{expt}A and corresponds to a threshold within the approximate range $10 \le x_* \le 25$. Therefore, we divide the $x$ axis into bins of size $\Delta x$, choose bin midpoints $x_*$ within this range, and for each choice compute the mean $t_*$ and the variance $\sigma_t^2$ of the data in that bin. Fig.\ \ref{expt}B (circles) shows the results for different values of $x_*$ (size) and $\Delta x$ (color).

The data in Fig.\ \ref{expt}B (circles) exhibit two clear features: (i) the dynamics are nonlinear ($\rho$ is significantly below $1$), and (ii) the timing variance is low ($\sigma^2_tx_*/t_*^2$ is significantly below $1$). Neither feature can be explained by a model in which the production of $x$ is unregulated, since that would correspond to a linear increase of molecule number over time ($\rho = 1$) and a timing variance that satisfies $\sigma^2_tx_*/t_*^2 = 1$ (square in Fig.\ \ref{expt}B). Furthermore, since auto-regulation has been shown to generically increase timing variance beyond the unregulated case \cite{ghusinga2017first}, it is unlikely that feature (ii) can be explained by a model with auto-regulation alone. Can these data be explained by our model with regulation?

To address this question we calculate $\rho$ and $\sigma^2_tx_*/t_*^2$ from our model. For simplicity, we focus on the analytic approximations in Eqs.\ \ref{varA} and \ref{varR}, since they have been validated in Fig.\ \ref{heatmaps}. In these approximations, since $\bar{x}(t)$ is piecewise-linear (Eq.\ \ref{xbar}), calculating $\rho$ via Eq.\ \ref{rho} is straightforward: $\rho = 1-t_0/t_*$, where $t_0$ is once again determined by either $\bar{a}(t_0) = K$ or $\bar{r}(t_0) = K$ according to Eqs.\ \ref{abar} and \ref{rbar}. For a given $\rho$ and cost $\langle a\rangle/x_*$ or $\langle r\rangle/x_*$, we calculate the minimum timing variance $\sigma^2_tx_*/t_*^2$. For the activator, we use the expression for $\rho$ along with Eq.\ \ref{costA} to write Eq.\ \ref{varA} in terms of $\rho$ and $\langle a\rangle/x_*$,
\begin{equation}
\label{rhoA}
\frac{\sigma^2_tx_*}{t_*^2} = \frac{x_*}{2\langle a\rangle}(1-\rho)+\rho^2.
\end{equation}
For the repressor, we use the expression for $\rho$ along with Eq.\ \ref{costR} to write Eq.\ \ref{varR} in terms of $\rho$ and $\langle r\rangle/x_*$, and then minimize over $N$ (see Materials and Methods) to obtain
\begin{equation}
\label{rhoR}
\frac{\sigma^2_tx_*}{t_*^2} = \frac{e^3}{27}\frac{x_*}{\langle r\rangle}(1-\rho)^3+\rho^2.
\end{equation}
Eqs.\ \ref{rhoA} and \ref{rhoR} are shown in Fig.\ \ref{expt}B (green solid and red dashed curves, respectively), and we see the same qualitative features for both cases: all curves satisfy $\sigma^2_tx_*/t_*^2 = 1$ at $\rho = 1$, as expected; and as $\rho$ decreases, each curve exhibits a minimum whose depth and location depend on cost. Specifically, as cost increases (lighter shades of green or red), the variance decreases, as expected. Importantly, we see that above a cost of $\langle a\rangle/x_* = \langle r\rangle/x_* \sim 10$, the model becomes consistent with the experimental data. This suggests that either an accumulating activator or a degrading repressor is sufficient to explain the temporal precision observed in {\it mig-1}-controlled neuroblast migration.

\subsection{Cell division implicates repression over activation}

The results in Fig.\ \ref{expt}B cannot distinguish between the two possibilities of an accumulating activator or a degrading repressor. However, there is a key feature of the data that is not yet accounted for in the model. Specifically, near the end of migration, cell division occurs (Fig.\ \ref{expt}A, black data). One daughter cell continues migrating (Fig.\ \ref{expt}A, dark blue data), while the other undergoes programmed cell death \cite{mentink2014cell}. To investigate the effects of cell division on the temporal precision of threshold crossing, we introduce cell division into the model. Specifically, we perform stochastic simulations \cite{gillespie1977exact} of the reactions in Fig.\ \ref{setup}A, and at a given time $t_d$, we reduce the molecule numbers of both the regulator and the target molecule. We assume symmetric partitioning of molecules, such that the molecule number after division is drawn from a binomial distribution with total number of trials equal to the molecule number before division, and success probability equal to one half. For each simulation, $t_d$ is drawn from a Gaussian distribution with mean $\bar{t}_d$ and variance $\sigma_d^2$ determined by the data (Fig.\ \ref{expt}A, black). As before, the maximal production rate $\alpha$ is set to ensure that the mean threshold crossing time $\bar{t}$ over all simulations equals $t_*$.

\begin{figure}
\begin{center}
\includegraphics[width=.5\textwidth]{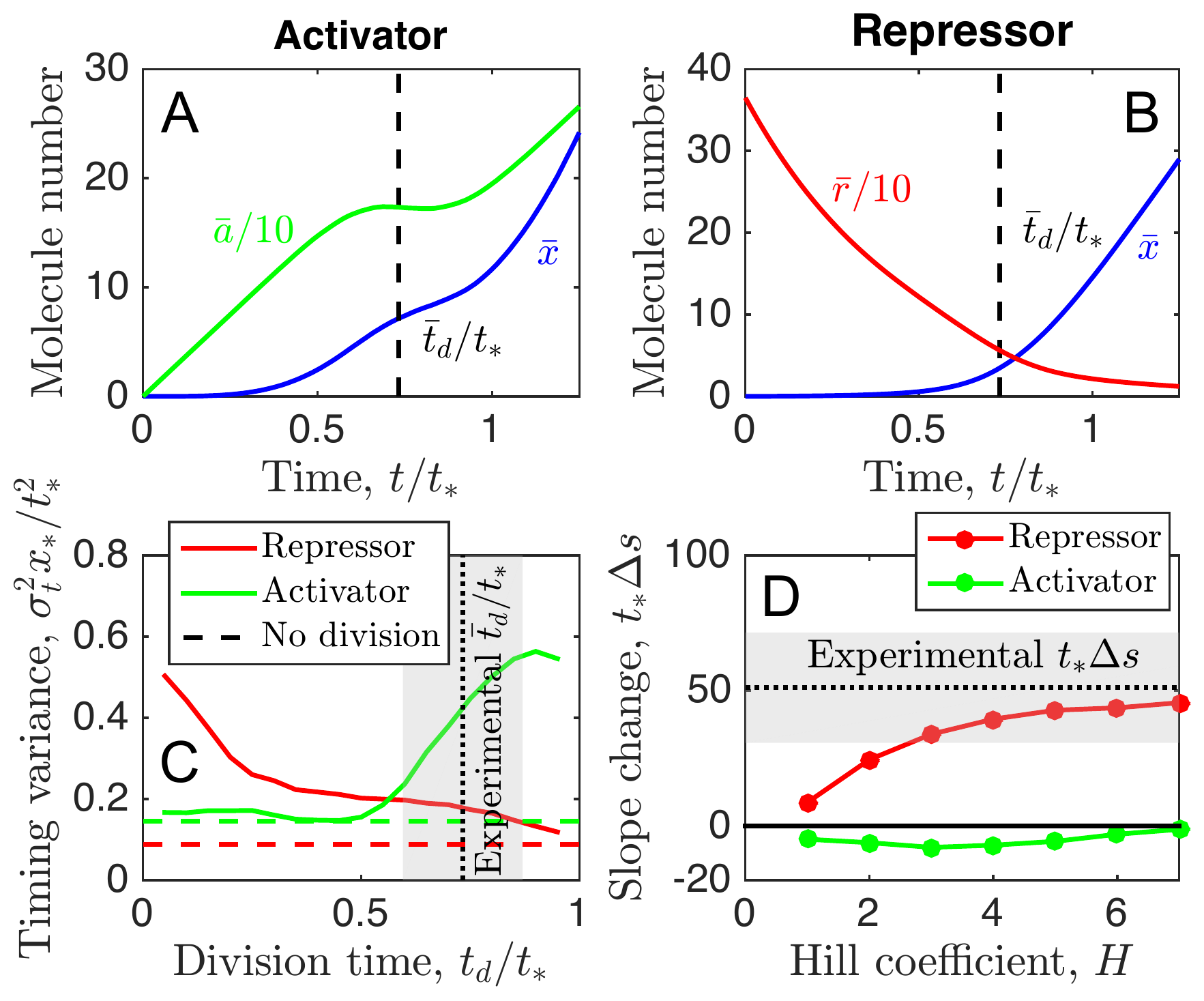}
\end{center}
\caption{Cell division implicates repression over activation. Mean dynamics from stochastic simulations of (A) activator model and (B) repressor model in which cell division occurs at time $\bar{t}_d$ on average. Abrupt reductions in molecule numbers are smoothed by noise in $t_d$ and by binomial partitioning of molecules. (C) Timing variance is lower when division is early for activator, or late for repressor, approaching that with no division (dashed). Division is late in experiments (gray). (D) Change in slope of target molecule dynamics after division is negative for activator, or positive for repressor. Change in slope is positive in experiments (gray) and agrees with repressor model for $H\gtrsim3$. Parameters are $x^*=15$, $\langle a\rangle/x_* = \langle r\rangle/x_* = 10$, $kt_*$ and $K$ set to optimal values (Fig.\ \ref{heatmaps}), $\bar{t}_d$ and $\sigma_d$ set to experimental values, and $H=3$.}
\label{division}
\end{figure}

Figure \ref{division}A and B show the average molecule numbers as a function of time for the activator and repressor cases, respectively. We see that the curves are affected by cell division, but that the abrupt reduction in molecule number is smoothed by the partitioning noise and the variability $\sigma_d^2$ in the times at which division occurs. Thus there is no abrupt reduction in molecule number, consistent with the data in Fig.\ \ref{expt}A. Any remaining reduction is even less pronounced in the repressor case because the molecule numbers at the time of division are lower than those in the activator case, and therefore they drop by less when reduced.

Figure \ref{division}C shows the scaled timing variance of threshold crossing $\sigma^2_tx_*/t_*^2$ as a function of division time $t_d$ in the model (solid curves). We see that even with the added noise of cell division, the timing variance is less than that of the unregulated case with no cell division ($\sigma^2_tx_*/t_*^2 = 1$). Furthermore, we see that if cell division occurs at early or late times for the activator or repressor case, respectively, then the added noise of cell division becomes small, and the timing variance approaches that of the case with no cell division (dashed lines). Intuitively, these regimes correspond to the times at which the regulator molecule number is low, and therefore, as discussed above, the effect of reducing the molecule number is less pronounced. In the experimental data, cell division occurs late in the process ($t_d/t_* > 0.5$, Fig. \ref{expt}A). In this regime the timing variance of the repressor case is lower than that of the activator case (Fig.\ \ref{division}C, gray region). This suggests that because neuroblasts in {\it C. elegans} larvae divide late in the migration process, they would benefit more from {\it mig-1} repression than {\it mig-1} activation for optimal temporal precision.

The results in Fig.\ \ref{division}C suggest that repression would produce higher temporal precision than activation, but they do not demonstrate that repression is actually occurring in the experiments. However, there is an additional feature of the data that may distinguish between activation and repression more directly. Specifically, we observe that the slope of the {\it mig-1} increase over time is steeper after cell division than before cell division (Fig.\ \ref{expt}A; see Materials and Methods). Therefore we investigate the equivalent slope change in the model. That is, we calculate the slopes $s_1 = d\bar{x}/dt$ before and $s_2 = d\bar{x}/dt$ after the mean division time (at $\bar{t}_d\pm\sigma_d$, as in the experiments). The difference $\Delta s = s_2 - s_1$ is shown in Fig.\ \ref{division}D as a function of the Hill coefficient $H$. We see that for activation $\Delta s$ is negative, whereas for repression $\Delta s$ is positive. The reason is that when the activator molecule number is reduced due to cell division, the production rate of the target molecule is also reduced, which reduces the slope of $\bar{x}$. In contrast, when the repressor molecule number is reduced, the production rate of the target molecule is increased, which increases the slope of $\bar{x}$. Thus, as shown in Fig.\ \ref{division}D, the increase in the slope of the experimental data after cell division is consistent with the repressor model, and in particular with cooperative repression ($H \gtrsim 3$), but not with the activator model. This result offers direct evidence that the regulation of {\it mig-1} is dominated by repression and not activation.

\section{Discussion}

We have demonstrated that regulation by an accumulating activator or a diminishing repressor increases the precision of threshold crossing by a target molecule, beyond the precision achievable with constitutive expression alone. The increase in precision results from a tradeoff between reducing the extrinsic noise of the regulator, and reducing the intrinsic noise of the target molecule itself. Our minimal model is sufficient to explain both the high degree of nonlinearity and the low degree of noise in the dynamics of {\it mig-1} in {\it C. elegans} neuroblasts, suggesting that these cells use regulated expression to terminate their migration with increased temporal precision. Moreover, the effects of cell division on the {\it mig-1} dynamics are consistent with our repressor model, but not our activator model, indicating that the regulation of {\it mig-1} is dominated by repression. These results suggest that regulation by a dynamic upstream species is a simple and generic method of increasing the temporal precision of cellular behaviors governed by threshold-crossing events.

Why does regulation increase temporal precision, whereas it has been shown that auto-regulation (feedback) does not \cite{ghusinga2017first}? After all, either regulation or positive feedback can produce an acceleration in molecule number over time, leading to a steeper threshold crossing. The reason is likely that positive feedback relies on self-amplification. In addition to amplifying the mean, positive feedback also amplifies the noise. In contrast, regulation by an external species does not involve self-amplification. Therefore, the noise increase is not as strong. The target molecule certainly inherits noise from the regulator (Eq.\ \ref{ie}), but the increase in noise does not outweigh the benefit of the acceleration, as it does for feedback.

Our finding that regulation increases temporal precision is related to the more basic phenomenon that a sequence of ordered events has a lower relative timing error than a single event. Specifically, as mentioned above, if a single event occurs in a time that is exponentially distributed with a mean $\mu$, then the relative timing error is $\sigma/\mu = 1$. However, for $n$ such events that must occur in sequence, the total completion time follows a gamma distribution with relative timing error $\sigma/\mu = 1/\sqrt{n}$, which decreases with increasing $n$. Thus, at a coarse-grained level, the addition of a regulator can be viewed as increasing the length of the sequence of threshold-crossing events from one to two, and thus decreasing the timing error. This perspective suggests that the timing error could be decreased even further via a cascade of regulators.

Our model neglects more complex features of regulated gene expression, such as bursts and degradation. Future work could investigate the interplay of production and degradation, or the interplay of regulation and feedback, especially as {\it mig-1} is thought to be subject to degradation and feedback in addition to external regulation \cite{ji2013feedback, mentink2014cell}. We anticipate that exploring the effects of the these features will lead to new fundamental insights about cellular timing precision, beyond the mechanisms elucidated here.

\section{Materials and methods}

\subsection{Computation of the first-passage time statistics}

We compute the first-passage time statistics $\bar{t}$ and $\sigma^2_t$ numerically from the master equation following \cite{ghusinga2017first}, generalized to a two-species system. Specifically, the probability $F(t)$ that the molecule number crosses the threshold $x_*$ at time $t$ is equal to the probability $P_{y,x_*-1}(t)$ that there are $y$ regulator molecules (where $y \in \{a,r\}$) and $x_*-1$ target molecules, and that a production reaction occurs with rate $f_\pm(y)$ to bring the target molecule number up to $x_*$. Since this event can occur for any regulator molecule number $y$, we sum over all $y$,
\begin{equation}
\label{fpt}
F(t) = \sum_{y=0}^Y f_\pm(y)P_{y,x_*-1}(t),
\end{equation}
where $Y = \{a_{\max},N\}$. The repressor has a maximum number of molecules $N$, whereas the activator number is unbounded, and therefore we introduce the numerical cutoff $a_{\max} = kt_* + \sqrt{10kt_*}$. The probability $P_{yx}$ evolves in time according to the master equation corresponding to the reactions in Fig.\ \ref{setup}A,
\begin{subequations}
\label{mboth}
\begin{align}
\label{ma}
\dot{P}_{ax} &= kP_{a-1,x} +f_+(a)P_{a,x-1} -[k+f_+(a)]P_{ax},\\
\label{mr}
\dot{P}_{rx} &= k(r+1)P_{r+1,x} +f_-(r)P_{r,x-1} -[kr + f_-(r)]P_{rx}.
\end{align}
\end{subequations}
The moments of Eq.\ \ref{fpt} are
\begin{equation}
\label{tm}
\langle t^m\rangle = \sum_{y=0}^Y f_\pm(y) \int_0^\infty dt\ t^mP_{y,x_*-1}(t),
\end{equation}
where $\bar{t} = \langle t\rangle$ and $\sigma_t^2 = \langle t^2\rangle - \langle t\rangle^2$. Therefore computing $\bar{t}$ and $\sigma^2_t$ requires solving for the dynamics of $P_{yx}$.

Because Eq.\ \ref{mboth} is linear in $P_{yx}$, it is straightforward to solve by matrix inversion. We rewrite $P_{yx}$ as a vector by concatenating its columns, $\vec{P}^\top = \left[[P_{00}, \dots, P_{Y0}], \dots, [P_{0,x_*-1}, \dots, P_{Y,x_*-1}]\right]$, such that Eq.\ \ref{mboth} becomes $\dot{\vec{P}} = {\bf M}\vec{P}$, where
\begin{equation}
\label{Mmat}
{\bf M} = \left[
\begin{matrix}
{\bf M}^{(1)} & & & & \\
{\bf M}^{(2)} & {\bf M}^{(1)} & & & \\
& {\bf M}^{(2)} & {\bf M}^{(1)} & & \\
& & \ddots & \ddots & \\
& & & {\bf M}^{(2)} & {\bf M}^{(1)} \\
\end{matrix}
\right].
\end{equation}
Here, for $i,j\in\{0,\dots,Y\}$, the $x_*-1$ subdiagonal blocks are the diagonal matrix ${\bf M}^{(2)}_{ij} = f_\pm(i)\delta_{ij}$, and the $x_*$ diagonal blocks are the subdiagonal matrix ${\bf M}^{(1)}_{ij} = -[k(1-\delta_{ia_{\max}})+f_+(i)]\delta_{ij}+k\delta_{i-1,j}$ or the superdiagonal matrix ${\bf M}^{(1)}_{ij} = -[ki+f_-(i)]\delta_{ij}+k(i+1)\delta_{i+1,j}$ for the activator or repressor case, respectively. The $\delta_{ia_{\max}}$ term prevents activator production beyond $a_{\max}$ molecules. The final ${\bf M}^{(1)}$ matrix in Eq.\ \ref{Mmat} contains $f_\pm$ production terms that are not balanced by equal and opposite terms anywhere in their columns. These terms correspond to the transition from $x_*-1$ to $x_*$ target molecules, for which there is no reverse transition. Therefore, the state with $x_*$ target molecules (and any number of regulator molecules) is an absorbing state that is outside the state space of $\vec{P}$ \cite{ghusinga2017first}. Consequently, probability leaks over time, and $\vec{P}(t\to\infty) = \vec{\emptyset}$. Equivalently, the eigenvalues of ${\bf M}$ are negative.

The solution of the dynamics above Eq.\ \ref{Mmat} is $\vec{P}(t) = \exp({\bf M}t)\vec{P}(0)$. Therefore, Eq.\ \ref{tm} becomes $\langle t^m\rangle = \vec{V}^\top \left[ \int_0^\infty dt\ t^m \exp({\bf M}t) \right] \vec{P}(0)$, where $\vec{V}^\top$ is a length-$x_*(Y+1)$ row vector consisting of $[f_\pm(0), \dots, f_\pm(Y)]$ preceded by zeros. We solve this equation via integration by parts \cite{ghusinga2017first}, noting that the boundary terms vanish because the eigenvalues of ${\bf M}$ are negative, to obtain
\begin{equation}
\langle t^m\rangle = (-1)^{m+1}m!\vec{V}^\top\left({\bf M}^{-1}\right)^{m+1}\vec{P}(0).
\end{equation}
We see that computing $\bar{t} = \langle t\rangle$ and $\sigma_t^2 = \langle t^2\rangle - \langle t\rangle^2$ requires inverting ${\bf M}$, which we do numerically in Matlab. We initialize $\vec{P}$ as $P_{ax}(0) = \delta_{a0}\delta_{x0}$ or $P_{rx}(0) = \delta_{rN}\delta_{x0}$ for the activator or repressor case, respectively.

When including cell division, we compute $\bar{t}$ and $\sigma^2_t$ from 50,000 stochastic simulations \cite{gillespie1977exact}.

\subsection{Deterministic dynamics}

The dynamics of the mean regulator and target molecule numbers are obtained by calculating the first moments of Eq.\ \ref{mboth}, $\partial_t\bar{y} = \sum_{yx} y\dot{P}_{yx}$ and $\partial_t\bar{x} = \sum_{yx} x\dot{P}_{yx}$, where $y \in \{a,r\}$. For the regulator we obtain $\partial_t\bar{a} = k$ or $\partial_t\bar{r} = -k\bar{r}$ in the activator or repressor case, respectively, which are solved by Eqs.\ \ref{abar} and \ref{rbar}. For the target molecule we obtain $\partial_t\bar{x} = \langle f_\pm(y)\rangle$, which is not solvable because $f_\pm$ is nonlinear (i.e., the moments do not close). A deterministic analysis conventionally assumes $\langle f_\pm(y)\rangle \approx f_\pm(\bar{y})$, for which the equation becomes solvable by separation of variables. For example, in the case of $H=1$, using Eqs.\ \ref{f+}-\ref{rbar}, we obtain
\begin{equation}
\label{det}
\bar{x}(t) = \frac{\alpha}{k}
\begin{cases}
kt - K\log \frac{kt+K}{K} & {\rm (activator)} \\
\log\frac{N+Ke^{kt}}{N+K} & {\rm (repressor)}.
\end{cases}
\end{equation}
Equation \ref{det} is plotted in Fig.\ \ref{setup}C and D.

When including cell division, we compute the mean dynamics from the simulation trajectories (Fig.\ \ref{division}A and B).

\subsection{Details of the analytic approximations}

To find the global minimum of Eq.\ \ref{varR}, we differentiate with respect to $kt_*$ and $K$ and set the results to zero, giving two equations. $kt_*$ can be eliminated, leaving one equation for $K$,
\begin{equation}
\frac{1}{2}\log\frac{N}{K} = 1 - \frac{K}{N}
\end{equation}
This equation is transcendental. However, in the limit $K\ll N$, we neglect the last term, which gives Eq.\ \ref{minR}.

To derive Eq.\ \ref{rhoR}, we use
\begin{equation}
\label{reprho}
\rho = 1 - \frac{t_0}{t_*} = 1 - \frac{\log N/K}{kt_*}
\end{equation}
where the second step follows from $\bar{r}(t_0) = K$ according to Eq.\ \ref{rbar}; and, from Eq.\ \ref{costR},
\begin{equation}
\label{tavgr}
\langle r \rangle = \frac{N}{kt_*}(1-e^{-kt_*}) \approx \frac{N}{kt_*}
\end{equation}
where the second step assumes that the repressor is fast-decaying, $kt_* \gg 1$.
We use Eqs.\ \ref{tavgr} and \ref{reprho} to eliminate $kt_*$ and $K$ from Eq.\ \ref{varR} in favor of $\rho$ and $\langle r\rangle$,
\begin{equation}
\label{varRrho}
\frac{\sigma^2_tx_*}{t_*^2} \approx \frac{x_*}{N}\Big(e^{N(1-\rho)/\langle r \rangle} -1\Big)\frac{\langle r \rangle ^2}{N^3} + \rho^2.
\end{equation}
For nonlinear dynamics ($\rho < 1$) we may safely neglect the $-1$ in Eq.\ \ref{varRrho}. Then, differentiating Eq.\ \ref{varRrho} with respect to $N$ and setting the result to zero, we obtain $N = 3\langle r \rangle/(1-\rho)$. Inserting this result into Eq.\ \ref{varRrho} produces Eq.\ \ref{rhoR}.

\subsection{Analysis of the experimental data}

To estimate the time at which migration terminates in Fig.\ \ref{expt}A, we refer to \cite{mentink2014cell}. There, the position at which neuroblast migration terminates is measured with respect to seam cells V1 to V6 in the larva (see Fig.\ 4D in \cite{mentink2014cell}). In particular, in wild type larvae, migration terminates between V2 and the midpoint of V2 and V1. This range corresponds to the magenta region in Fig.\ \ref{expt}A (see Fig.\ 4B, upper left panel, in \cite{mentink2014cell}). Under the assumptions of constant migration speed and equal distance between seam cells, the horizontal axis in Fig.\ \ref{expt}A represents time.

To compute $\rho$ for the experimental data in Fig.\ \ref{expt}A according to Eq.\ \ref{rho} we use a trapezoidal sum. For the choices of $x_*$ and $t_*$ described in the text, this produces the $\rho$ values in Fig.\ \ref{expt}B.

To estimate the slopes before and after cell division in Fig.\ \ref{expt}A, we use the following procedure. We perform a linear fit using the $D$ data points with times closest to $\bar{t}_d-\sigma_d$ (before division, light blue) or $\bar{t}_d+\sigma_d$ (after division, dark blue). We find that $5 \le D \le 15$ gives reasonable results, and we compute the mean and standard deviation of the slopes in this range. The slope difference, normalized by $t_*$, with error propagated in quadrature, is shown in Fig.\ \ref{division}D. Example fits with $D=8$ are shown in Fig.\ \ref{expt}A.

\section{Acknowledgments}

We thank Jeroen van Zon and Pieter Rein ten Wolde for helpful discussions, and van Zon for preliminary simulations that inspired the work. This work was supported by Human Frontier Science Program grant RGP0030/2016.


\end{document}